\documentclass{aa} 
\usepackage{txfonts}
\usepackage{longtable}  
\usepackage{rotating}
\usepackage{natbib}
\usepackage{graphicx}
\usepackage{graphics}
\usepackage{psfrag}
\usepackage{amssymb}
\bibpunct{(}{)}{;}{a}{}{,}
\def\Teff{$T_{\mathrm{eff}}$}
\def\logg{\ensuremath{\log g}}
\def\vmic{$\upsilon_{\mathrm{mic}}$}
\def\vmac{$\upsilon_{\mathrm{macro}}$}
\def\vsini{\ensuremath{{\upsilon}\sin i}}
\def\kms{$\mathrm{km\,s}^{-1}$}
\def\ms{$\mathrm{m\,s}^{-1}$}
\def\exc{$\chi_{\mathrm{excit}}$}
\def\loggf{log $gf$}

\def\nlte{non-LTE}
\def\llm{{\sc LLmodels}}
\def\hfs{{\it hfs}}

\def\width{{\sc WIDTH9}}
\def\synth{{\sc SYNTH3}}
\def\vald{{\sc VALD}}

\begin{document} 
\title{Improved fundamental parameters and LTE abundances of the CoRoT
\thanks{The CoRoT space mission was developed and is operated by the French 
space agency CNES, with participation of ESA's RSSD and Science Programmes, 
Austria, Belgium, Brazil, Germany, and Spain}
solar-type pulsator HD~49933} 
\subtitle{} 
\author{T. Ryabchikova\inst{1,2} \and
        L. Fossati\inst{1}       \and 
	D. Shulyak\inst{1}
}
\institute{
	Institut f\"ur Astronomie, Universit\"{a}t Wien, 
	T\"{u}rkenschanzstrasse 17, 1180 Wien, Austria.\\
	\email{ryabchik@astro.univie.ac.at,fossati@astro.univie.ac.at,denis.shulyak@gmail.com} 
	\and
	Institute of Astronomy, Russian Academy of Sciences, Pyatnitskaya 
	48, 119017 Moscow, Russia.
	\email{ryabchik@inasan.ru}
} 
\date{} 
\abstract
{}
{The knowledge of accurate stellar parameters is a key stone in several fields
of stellar astrophysics, such as asteroseismology and stellar evolution. 
Although the parameters can be derived both via spectroscopy and multicolor 
photometry, the obtained results are sometimes affected by systematic 
uncertainties. In this paper we present a self-consistent spectral 
analysis of the solar-type star HD~49933, which is a primary target for 
the CoRoT satellite.}
{We used high-resolution and high signal-to-noise ratio spectra to carry 
out a consistent parameter estimation and abundance analysis of HD~49933. 
The \llm\ code was employed for model atmosphere calculations, while \synth\ 
and \width\ codes were used for line profile calculation and LTE abundance 
analysis.}
{In this paper we provide a detailed description of the methodology 
adopted to derive the fundamental parameters and the abundances. Although the 
obtained parameters differ from the ones previously derived by other authors, 
we show that only the set obtained in this work is able to fit the 
observed spectrum accurately. In particular, the surface gravity was adjusted 
to fit pressure-sensitive spectral features.}
{We confirm the importance of a consistent analysis of relevant spectroscopic 
features, application of advanced model atmospheres, and the use of up-to-date 
atomic line data for the determination of stellar parameters. These 
results are crucial for further studies, e.g. detailed theoretical modelling 
of the observed pulsation frequencies.}
\titlerunning{}
\authorrunning{T.~Ryabchikova et al.}
\maketitle
\section{Introduction}\label{introduction}
In the last ten years several space missions were launched to obtain high 
quality photometric data (e.g. WIRE, MOST, CoRoT) and several others are going 
to be launched in the near future (e.g. KEPLER, BRITE). One of the main 
goals of these missions is to provide astronomers with high precision 
photometric data that will allow a better understanding of stellar 
pulsation -- the only approach that can improve our knowledge about 
stellar interiors.

The modelling of pulsational signals requires the knowledge of stellar
parameters and primarily accurate values of the effective temperature (\Teff) 
and metallicity ($Z$). The determination of fundamental parameters can 
be performed by different methods
\citep[some examples for stars from B- to G-type are:][]{fuhrmann,przybilla2006,fossati2009} 
that not always lead to consistent results. Thus it is important to 
choose a methodology that allows to constrain the parameters of the star from 
the available observables (usually photometry and spectroscopy) in the most 
robust and reliable way.

After \citet{mosser2005} discovered the presence of solar-like 
oscillations in HD~49933 using RV measurements from high resolution
time-series spectra, this star was included as a primary CoRoT target for the 
photometric observation of solar-like oscillations. It was
one of the first main sequence solar-type star in which solar-like
oscillations were clearly detected from space photometry, 
thanks to the high quality data provided by the CoRoT satellite \citep{app}. 

Fundamental parameters and abundances of HD~49933 were obtained previously 
by different authors with different methods. The published fundamental 
parameters of HD~49933 display some scatter both in \Teff\ and \logg\ 
\citep[e.g. ][]{edvardsson,blackwell,gillon,cenarro,bruntt}. As a 
consequence, the published abundances show some scatter.
At the same time a need for a higher precision in stellar parameters for 
HD~49933 was emphasised by recent asteroseismic modelling of this object 
\citep{app,thomas}, which employed new pulsation analysis techniques
\citep[e.g. ][]{gruberbauer}.  

The main goal of the present work is to perform a consistent atmospheric 
and abundance analysis of HD~49933 which fits all photometric and 
spectroscopic data and to investigate to what degree the derived 
fundamental parameters depend on the applied methods.

\section{Observations and spectral reduction}\label{observations}
Ten spectra of HD~49933 were obtained between February 11th and February 13th
2006 with the cross-dispersed echelle spectrograph HARPS (spectral
resolution $R$ $\sim$ 115\,000) attached to the 3.6-m ESO La Silla 
telescope. The spectra were reduced with the online pipeline\footnote{\tt
http://www.eso.org/sci/facilities/lasilla/\\instruments/harps/tools/software.html}.

Each spectrum was obtained with an exposure time between 243 and 300 seconds. 
We retrieved the spectra from the ESO archive and co-added them to obtain a 
single spectrum with a signal-to-noise ratio (SNR) per pixel of about 
500, calculated at~$\sim$5000\,\AA. The final spectrum, normalised by 
fitting a low order spline to carefully selected continuum points, covers the 
wavelength range 3780--6910\,\AA. The spectrum has a gap between 5300\,\AA\ and 
5330\,\AA, since one echelle order is lost in the gap 
between the two chips of the CCD mosaic detector. 

\begin{figure}[ht]
\begin{center}
\includegraphics[width=\hsize,clip]{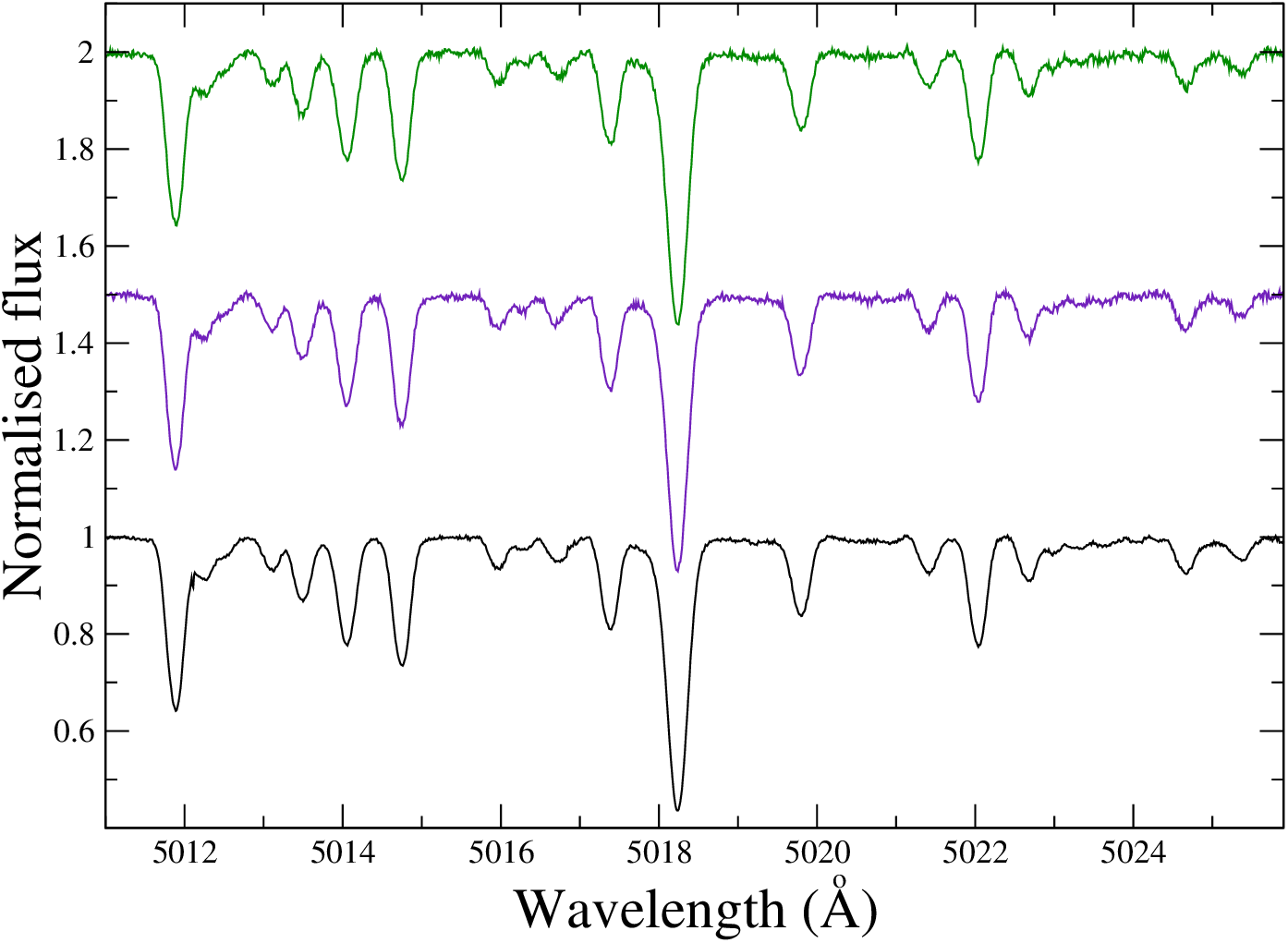}
\caption{Samples of the first spectrum obtained February 11th, the last 
obtained February 13th and the final co-added spectrum (from top to bottom)
around the strong \ion{Fe}{ii} line at $\sim$5018\,\AA.}
\label{portions}
\end{center}
\end{figure}

Figure~\ref{portions} (Online material) shows (from top to bottom) a sample of 
the first spectrum obtained on February 11th, the last obtained on February 
13th and the final co-added spectrum, around the strong \ion{Fe}{ii} 
line at 5018\,\AA. 

\section{The model atmosphere}
To compute model atmospheres of HD~49933 we employed the \llm\ stellar 
model atmosphere code \citep{llm}. For all the calculations Local 
Thermodynamical Equilibrium (LTE) and plane-parallel geometry were assumed. 
We used the \vald\ database \citep{vald1,vald2,vald3} as a source of atomic 
line parameters for opacity calculations. For a given model atmosphere 
we performed a line selection procedure which allowed to chose lines 
that significantly contribute to the opacity for a given set of model 
parameters, adopting the selection threshold 
$\ell_{\rm\nu}/\alpha_{\rm\nu} \geqslant 1$\%, where $\alpha_{\rm\nu}$ and 
$\ell_{\rm\nu}$ are the continuum and line absorption coefficients at a 
given frequency $\rm\nu$. 
Convection was implemented according to the \citet{cm1,cm2} model of 
convection \citep[see ][ for more details]{heiter}. 

\section{Fundamental parameters and abundance analysis}\label{parameters}
\citet{gillon} used Str\"omgren indices to determine the atmospheric 
parameters of HD~49933. By using the TEMPLOGG code \citep{templogg} they 
found \Teff\ = 6543$\pm$200\,K, \logg\ = 4.24$\pm$0.20 and 
${\rm [Fe/H]} = -0.38\pm0.20$\,dex. Although the uncertainties are quite 
large, we used this as our starting point and in an iterative process to 
gradually improve the parameters using different spectroscopic indicators 
as we will describe in detail below. In our analysis, every time any 
of \Teff, \logg, \vmic\ or abundances changed during the iteration process, 
we recalculated a new model with the implementation of the last 
measured quantities. The same concerns newly derived abundances: 
while the results of the abundance analysis depend upon the assumed model 
atmosphere, the atmospheric temperature-pressure structure itself depends 
upon the adopted abundances, so that we recalculated the model atmosphere 
every time abundances were changed, even if the other model parameters were 
fixed. Such a procedure ensures the model structure to be consistent with the 
assumed abundances.

We performed the \Teff\ determination by fitting synthetic
line profiles, calculated with \synth\ \citep{synth3}, to the observed profiles
of three hydrogen lines: H$\alpha$, H$\beta$ and H$\gamma$. In the temperature 
range expected for HD~49933 hydrogen lines are extremely sensitive to 
temperature variations and very little to \logg\ variations, therefore being 
good temperature indicators. The \Teff\ obtained with this procedure is 
\Teff\ = 6500 $\pm$ 50\,K.

\begin{figure}[ht]
\begin{center}
\includegraphics[width=\hsize,clip]{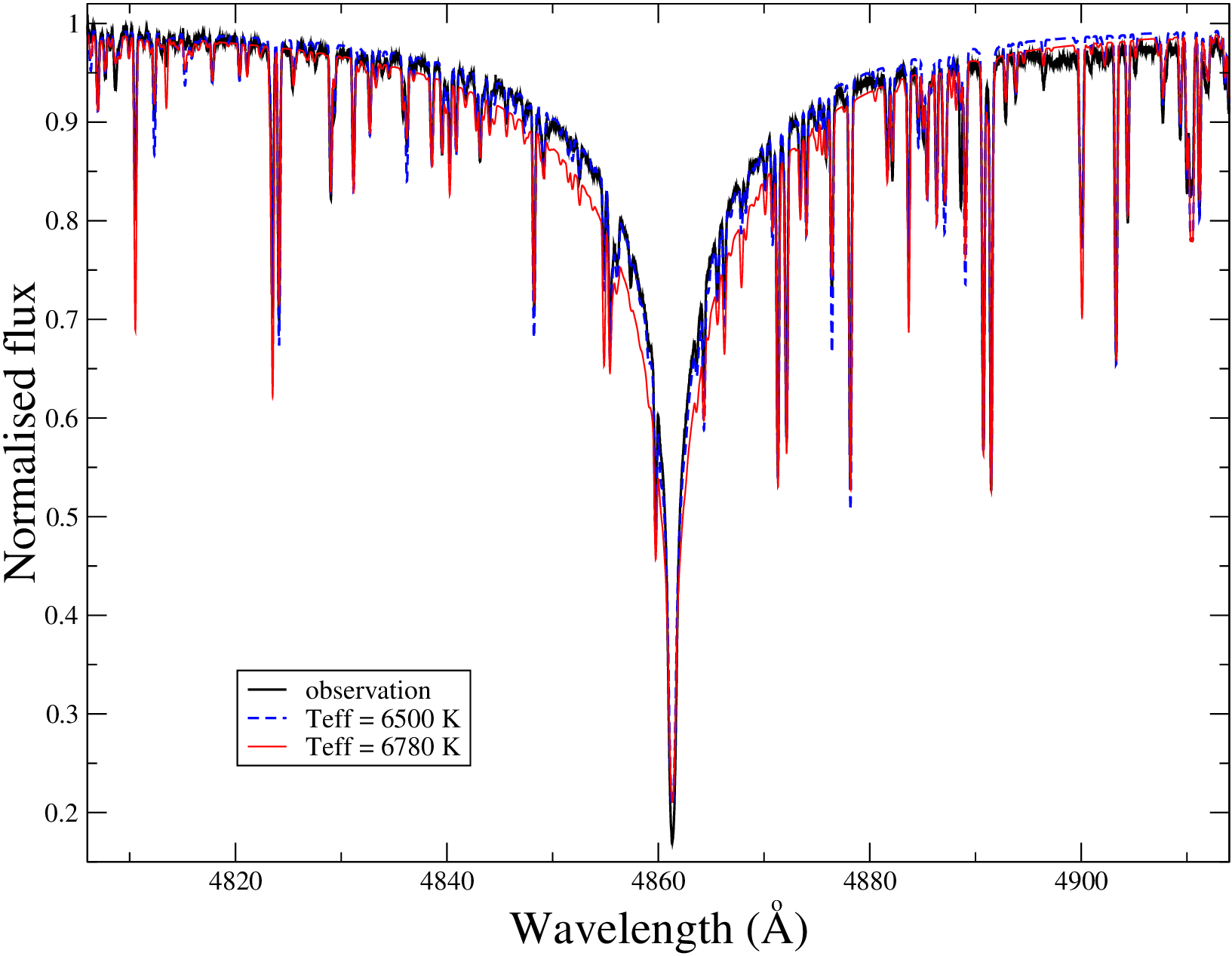}
\caption{Comparison between the observed H$\beta$ line profile (solid line) and
synthetic profiles calculated with the final adopted \Teff$=6500$\,K (dashed
line) and with the \Teff=$6780$\,K from \citet{bruntt}  
(solid thin line). The dashed line agrees almost perfectly with the 
observed spectrum.}
\label{hydrogen}
\end{center}
\end{figure}

Figure~\ref{hydrogen} shows the comparison between the observed H$\beta$ line 
profile and the synthetic profiles calculated with the adopted stellar 
parameters, as well as the synthetic profile obtained with the higher
\Teff\ = 6780 $\pm$ 130\,K published by \citet{bruntt}.

Another spectroscopic indicator for \Teff\ is given by the analysis of metallic
lines. In particular, \Teff\ is determined eliminating the correlation 
between line abundance and the line excitation potential (\exc) for a given 
ion/element. This procedure can lead to erroneous parameters, so we 
decided not to take into account this indicator in our analysis. This 
important point will be extensively discussed in Sect.~\ref{discussion}.

\begin{figure}
\begin{center}
\includegraphics[width=\hsize,clip]{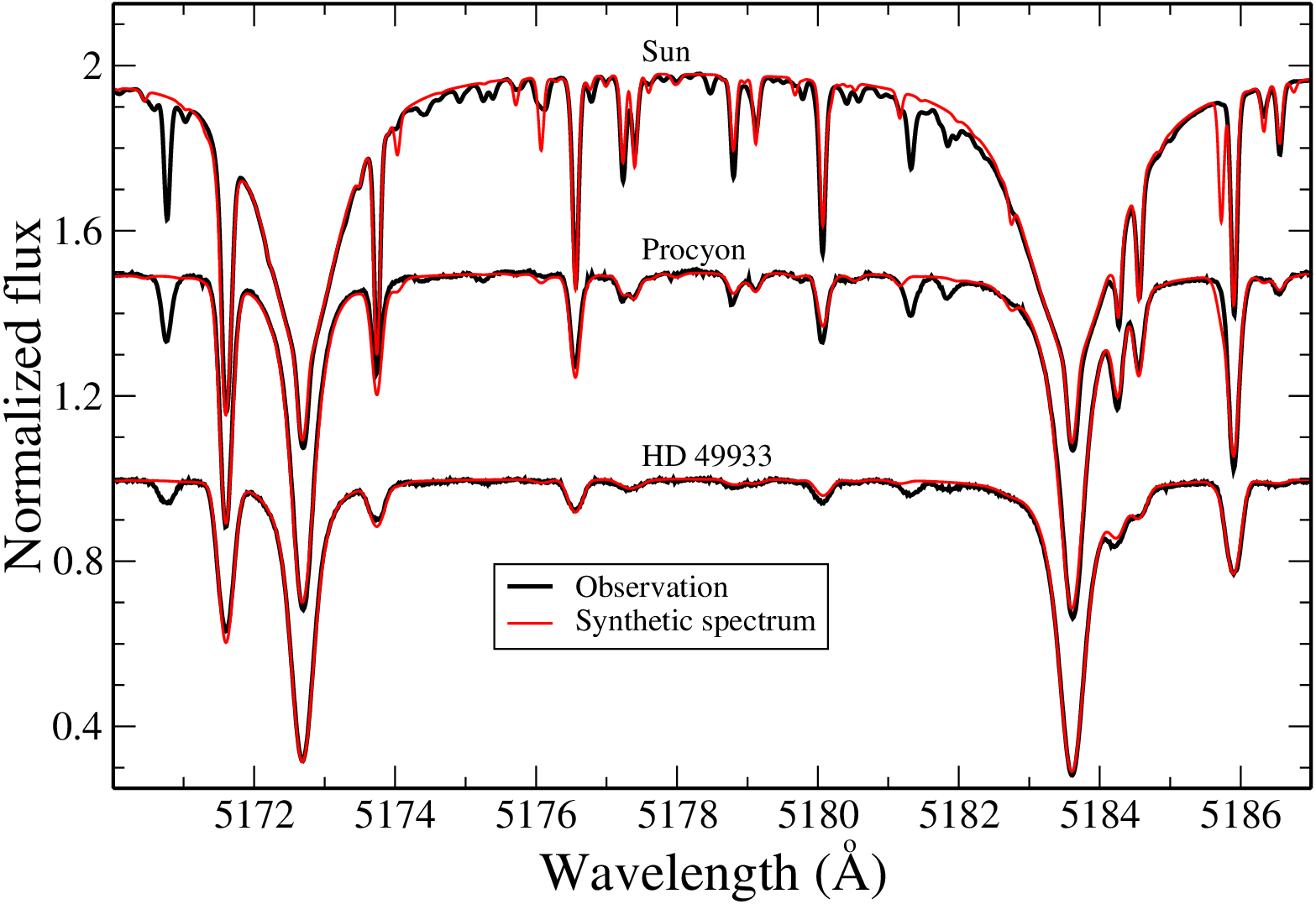}
\caption{Comparison between the observed \ion{Mg}{i} line profiles used to 
determine \logg\ (thick line) and synthetic profiles for the Sun, Procyon 
and HD~49933, from top to bottom. The synthetic spectrum shown for HD~49933 
was calculated with the final adopted \logg$=4.00$ (thin line). For all three
stars we adopted the same combination of \loggf\ and Van der Waals
damping constants: oscillator strengths from \citet{Aldenius} and damping 
constants from \citet{fuhrmann}. See text for details on the observed spectra 
and adopted model atmospheres for the Sun and Procyon. The spectra of the Sun 
and Procyon were shifted upwards by 1 and 0.5 respectively.}
\label{mglines}
\end{center}
\end{figure}

The surface gravity was derived from two independent methods based on 
line profile fitting of \ion{Mg}{i} lines with developed wings and ionisation 
balance for several elements. The first method is described in 
\citet{fuhrmann} and is based on the fact that the wings of the \ion{Mg}{i} 
lines at $\lambda\lambda$ 5167, 5172 and 5183\,\AA\ are very sensitive to 
\logg\ variations. In practice we derived first the Mg abundance from other 
\ion{Mg}{i} lines without developed wings, such as $\lambda\lambda$ 4571, 
4730, 5528 and 5711\,\AA, and then we fit the wings of the other three  
lines by tuning the \logg\ value. To apply this method, it is necessary 
to have very accurate \loggf\ values and Van der Waals 
(log\,$\mathbf{\gamma_{\rm Waals}}$) damping constants for all the lines. The 
\loggf\ values of the Mg lines given in the VALD database are of rather 
high quality and came from the laboratory measurements of the lifetimes 
\citep{AZ}. However, recently new laboratory measurements of the oscillator 
strengths for \ion{Mg}{i} lines at $\lambda\lambda$ 5167, 5172 and 5183\,\AA\ 
with an accuracy $\sigma$\loggf$=\pm$0.04\,dex were published by 
\citet{Aldenius}. Van der Waals damping constants in \vald\ are calculated by 
\citet{barklem} but they appear to be slightly higher than needed to 
fit the solar lines. While log\,$\mathbf{\gamma_{\rm Waals}}=-7.27$ in 
\citet{barklem}, \citet{fuhrmann} derived 
log\,$\mathbf{\gamma_{\rm Waals}}=-7.42$ from the fitting of the solar lines 
using the \citet{AZ} oscillator strengths. For our analysis we employed 
different combinations of oscillator strengths and damping constants. 
Adopting oscillator strengths from \citet{Aldenius} and damping constants 
from \citet{barklem} or oscillator strengths from \citet{AZ} and damping 
constants from \citet{fuhrmann} we derived a \logg\ value of 3.85, while
adopting oscillator strengths from \citet{Aldenius} and damping constants 
from \citet{fuhrmann} we obtained a \logg\ value of 4.00.
We obtained a formal \logg\ value of 3.93 $\pm$ 0.07 where the given error 
bar on \logg\ mainly depends on the uncertainty of the damping constant 
($\mathbf{\gamma_{\rm Waals}}$). Comparison of the synthetic colors with the 
observed ones (see Sect.~\ref{synth-phot}) favors a \logg\ value of 
$4.00\pm0.15$. Figure~\ref{mglines} shows a comparison between the synthetic 
and observed \ion{Mg}{i} line profile for the 5172 and 5183\,\AA\ \ion{Mg}{i} 
lines. The same procedure applied to the solar flux spectrum \citep{NSO} 
resulted in \logg=4.44 (oscillator strengths from Aldenius et al. 2007 
and damping constants from Fuhrmann et al. 1997) and \logg=4.30 for the 
combination of the oscillator strengths from \citet{Aldenius} and damping 
constants from \citet{barklem}. The comparison of the
observed and the synthetic \ion{Mg}{i} line profile for the 5172 and 
5183\,\AA\ \ion{Mg}{i} lines for the Sun is shown in Fig.~\ref{mglines}.  

For surface gravity determination one often uses ionisation equilibrium, 
but this method is extremely sensitive to the \nlte\ effects present for each 
ion/element, while Mg lines with developed wings (less sensitive to 
\nlte\ effects) are more suitable as \logg\ indicators \citep{zhao}. For this 
reason we decided to keep the Mg lines as our primary \logg\ indicator 
checking afterwards the obtained result with the ionisation equilibrium. The 
\logg\ value we obtained (\logg=4.00$\pm$0.15) is lower than that given 
by photometry and by other authors, e.g. \citet{gillon}:
\logg=4.26$\pm$0.08 and \citet{bruntt}: \logg=4.24$\pm$0.13. The difference 
between our result and, for example, the one of \citet{gillon,bruntt} is 
clearly connected with the difference in the derived effective 
temperature, while there is a reasonable agreement for \logg, but the 
error bars are unfortunately quite large. This important point will be 
discussed in Sect.~\ref{discussion}.

Since HD~49933 does not have an effective temperature high enough to show
He lines, we are not able to measure the atmospheric He abundance. Diffusion 
calculations for low-mass metal poor stars show that helium should be depleted 
in the atmospheric layers \citep{michaud}, therefore we tested the effect of 
a helium underabundance on the obtained parameters. We rederived \Teff\ and 
\logg\ respectively with hydrogen lines and \ion{Mg}{i} line profiles assuming 
a helium abundance of $-2.0$\,dex and of $-4.0$\,dex. In both cases we 
obtained no visible change of the hydrogen line profiles indicating no effect 
on the \Teff\ determination, while \logg\ should be increased of about 0.15 
with both helium underabundances,  to be able to fit again the 
\ion{Mg}{i} line profiles adopted to derive \logg. This value is within the 
adopted uncertainty for \logg.

Our main source for the atomic parameters of spectral lines is the \vald\ 
database. LTE abundance analysis was based on equivalent widths, analysed 
with a modified version \citep{vadim} of the WIDTH9 code \citep{kurucz1993a}.
Although 301 \ion{Fe}{i} lines were measured, only 158 lines with accurate 
experimental oscillator strengths were chosen for abundance determination
to achieve the highest accuracy. First, we reject the lines  with  
theoretical oscillator strengths. The rest of the lines were checked in the 
solar flux spectrum \citep{NSO}, observed with a resolving power 
$R \simeq 340\,000$ at wavelengths between 4000\,\AA\ and 4700\,\AA\ and 
with $R = 520\,000$ at longer wavelengths. The final choice was made for the 
lines that did not require an oscillator strength correction greater than 
0.1\,dex to fit the line cores of the solar spectrum. We also tried to have 
a set of \ion{Fe}{i} lines uniformly distributed over the range of equivalent 
widths and excitation potentials. As for the lines of other elements/ions we 
used nearly all unblended  spectral lines with accurate atomic 
parameters available in the  wavelength range 3850--6880\,\AA, except 
lines in spectral regions where the continuum normalisation was too uncertain. 
In case of blended lines, for lines subjected to hyperfine splitting (\hfs) or 
for lines situated in the wings of the hydrogen lines we derived the line 
abundance performing synthetic spectrum calculations with the \synth\ code. 
The \hfs\ constants for abundance calculations were taken from 
\citet{Mn1hfs05} for \ion{Mn}{i} lines, from \citet{Biehl1976} for 
\ion{Ba}{ii} $\lambda$~4554\,\AA\ line and from \citet{Eu2} for \ion{Eu}{ii} 
lines. \hfs\ effects are negligible for the two \ion{Cu}{i} lines we 
used  in the abundance analysis.
\begin{table*}[ht]
\caption[ ]{LTE atmospheric abundances HD~49933 with error
estimates based on the internal scatter from the number of analysed lines,
$n$. Fourth and fifth columns give HD~49933 abundances relative to the solar values from
\citet{met05} (AGS) and from \citet{sun_GS} (GS), respectively. The sixth column gives abundances derived
by Bruntt et al. (2008) relative to GS. The last column gives the abundances of the solar atmosphere
from \citet{met05}.}
\label{abundance}
\begin{center}
\begin{tabular}{l|cc|cc|c|c}
\hline
\hline
Ion &\multicolumn{4}{|c|}{HD~49933 - this paper}& Bruntt et al. (2008)  &  Sun \\                    
    &$\log (N/N_{\rm tot})$ & $n$  &[$N_{\rm el}/N_{\rm H}$]$_{\rm AGS}$ &[$N_{\rm el}/N_{\rm H}$]$_{\rm GS}$ &[$N_{\rm el}/N_{\rm H}$]$_{\rm GS}$ &$\log (N/N_{\rm tot})$  \\
\hline                                                                                                        
\ion{C}{i}    & ~~$-$3.74$\pm$0.10 &  6 & ~~$-$0.09& ~~$-$0.22 & ~~$-$0.56 & ~~$-$3.65~ \\			    
\ion{O}{i}    & ~~$-$3.55          &  1 & ~~$-$0.17& ~~$-$0.34 & ~~$-$0.53 & ~~$-$3.38~ \\			    
\ion{Na}{i}   & ~~$-$6.15$\pm$0.05 &  5 & ~~$-$0.28& ~~$-$0.44 & ~~$-$0.36 & ~~$-$5.87~ \\			    
\ion{Mg}{i}   & ~~$-$4.83$\pm$0.07 &  4 & ~~$-$0.32& ~~$-$0.37 &           & ~~$-$4.51~ \\			    
\ion{Mg}{ii}  & ~~$-$4.73          &  1 & ~~$-$0.22& ~~$-$0.27 &           & ~~$-$4.51~ \\			    
\ion{Al}{i}   & ~~$-$6.20:         &  2 & ~~$-$0.53& ~~$-$0.63 &           & ~~$-$5.67~ \\			    
\ion{Si}{i}   & ~~$-$4.86$\pm$0.21 & 20 & ~~$-$0.33& ~~$-$0.37 & ~~$-$0.37 & ~~$-$4.53~ \\			    
\ion{Si}{ii}  & ~~$-$4.82$\pm$0.02 &  6 & ~~$-$0.29& ~~$-$0.33 &           & ~~$-$4.53~ \\			    
\ion{S}{i}    & ~~$-$5.23$\pm$0.07 &  2 & ~~$-$0.33& ~~$-$0.52 & ~~$-$0.36 & ~~$-$4.90~ \\			    
\ion{Ca}{i}   & ~~$-$6.01$\pm$0.11 & 26 & ~~$-$0.28& ~~$-$0.33 & ~~$-$0.50 & ~~$-$5.73~ \\			    
\ion{Ca}{ii}  & ~~$-$6.01$\pm$0.09 &  9 & ~~$-$0.28& ~~$-$0.33 &           & ~~$-$5.73~ \\			    
\ion{Sc}{ii}  & ~~$-$9.24$\pm$0.12 & 12 & ~~$-$0.25& ~~$-$0.37 & ~~$-$0.45 & ~~$-$8.99~ \\			    
\ion{Ti}{i}   & ~~$-$7.54$\pm$0.07 & 19 & ~~$-$0.40& ~~$-$0.52 & ~~$-$0.52 & ~~$-$7.14~ \\			    
\ion{Ti}{ii}  & ~~$-$7.42$\pm$0.12 & 33 & ~~$-$0.28& ~~$-$0.40 & ~~$-$0.41 & ~~$-$7.14~ \\			    
\ion{V}{i}    & ~~$-$8.50$\pm$0.13 &  4 & ~~$-$0.46& ~~$-$0.46 &           & ~~$-$8.04~ \\			    
\ion{V}{ii}   & ~~$-$8.47$\pm$0.23 &  5 & ~~$-$0.43& ~~$-$0.43 &           & ~~$-$8.04~ \\			    
\ion{Cr}{i}   & ~~$-$6.82$\pm$0.17 & 25 & ~~$-$0.42& ~~$-$0.45 & ~~$-$0.63 & ~~$-$6.40~ \\			    
\ion{Cr}{ii}  & ~~$-$6.61$\pm$0.17 & 16 & ~~$-$0.21& ~~$-$0.24 & ~~$-$0.43 & ~~$-$6.40~ \\			    
\ion{Mn}{i}   & ~~$-$7.33$\pm$0.14 & 14 & ~~$-$0.68& ~~$-$0.68 &           & ~~$-$6.65~ \\			    
\ion{Fe}{i}   & ~~$-$5.04$\pm$0.06 &158 & ~~$-$0.45& ~~$-$0.50 & ~~$-$0.44 & ~~$-$4.59~ \\			    
\ion{Fe}{ii}  & ~~$-$5.03$\pm$0.08 & 31 & ~~$-$0.44& ~~$-$0.49 & ~~$-$0.44 & ~~$-$4.59~ \\			    
\ion{Co}{i}   & ~~$-$7.49$\pm$0.10 &  3 & ~~$-$0.37& ~~$-$0.37 &           & ~~$-$7.12~ \\			    
\ion{Ni}{i}   & ~~$-$6.34$\pm$0.10 & 41 & ~~$-$0.53& ~~$-$0.55 & ~~$-$0.48 & ~~$-$5.81~ \\			    
\ion{Cu}{i}   & ~~$-$8.65$\pm$0.07 &  2 & ~~$-$0.82& ~~$-$0.82 &           & ~~$-$7.83~ \\			    
\ion{Zn}{i}   & ~~$-$8.12$\pm$0.06 &  2 & ~~$-$0.66& ~~$-$0.66 &           & ~~$-$7.44~ \\			    
\ion{Sr}{i}   & ~~$-$9.65          &  1 & ~~$-$0.53& ~~$-$0.58 &           & ~~$-$9.12~ \\			    
\ion{Sr}{ii}  & ~~$-$9.50$\pm$0.04 &  3 & ~~$-$0.38& ~~$-$0.43 &           & ~~$-$9.12~ \\			    
\ion{Y}{ii}   & ~$-$10.34$\pm$0.10 &  5 & ~~$-$0.51& ~~$-$0.54 &           & ~~$-$9.83~ \\			    
\ion{Zr}{ii}  & ~~$-$9.85$\pm$0.06 &  5 & ~~$-$0.40& ~~$-$0.41 &           & ~~$-$9.45~ \\			    
\ion{Ba}{ii}  & ~$-$10.06$\pm$0.19 &  5 & ~~$-$0.19& ~~$-$0.15 &           & ~~$-$9.87~ \\			    
\ion{La}{ii}  & ~$-$11.21$\pm$0.11 &  5 & ~~$-$0.30& ~~$-$0.34 &           & ~$-$10.91~ \\			    
\ion{Ce}{ii}  & ~$-$10.73$\pm$0.10 &  5 & ~~$-$0.27& ~~$-$0.27 &           & ~$-$10.46~ \\			    
\ion{Nd}{ii}  & ~$-$10.77$\pm$0.28 &  8 & ~~$-$0.18& ~~$-$0.23 &           & ~$-$10.59~ \\			    
\ion{Sm}{ii}  & ~$-$11.09$\pm$0.16 &  3 & ~~$-$0.06& ~~$-$0.06 &           & ~$-$11.03~ \\			    
\ion{Eu}{ii}  & ~$-$11.92$\pm$0.10 &  2 & ~~$-$0.40& ~~$-$0.39 &           & ~$-$11.52~ \\			    
\ion{Gd}{ii}  & ~$-$11.16$\pm$0.09 &  4 & ~~$-$0.24& ~~$-$0.24 &           & ~$-$10.92~ \\			    
\ion{Dy}{ii}  & ~$-$11.36$\pm$0.15 &  2 & ~~$-$0.46& ~~$-$0.46 &           & ~$-$10.90~ \\			    
\hline											     %
\Teff     &\multicolumn{4}{|c|}{6500~K}    & 6780~K    & 5777~K  \\				
\logg     &\multicolumn{4}{|c|}{4.00~~~}   & 4.24~~~   & 4.44~~~~\\				 
\hline											  
\end{tabular}
\end{center}
\end{table*}


The projected rotational velocity and macroturbulence (\vmac)  
were  determined fitting synthetic spectra of several 
carefully selected lines in the observed spectrum. We obtained 
\vsini\ = 10 $\pm$ 0.5\,\kms\ and  \vmac\ = 5.2 $\pm$ 0.5\,\kms. The value 
derived for \vmac\ is in agreement with that expected according to the 
relation \Teff\ - \vmac\ published by \citet{valenti} and obtained on a 
sample of more than a thousand stars. 

\citet{mosser2005} detected a line profile distortion variable with time with 
an amplitude of less than 500\,\ms, probably due to granulation. This 
distortion is too small to be able to affect the 
parameter and abundance determination since the parameters were mainly 
derived with hydrogen and \ion{Mg}{i} line profiles, while abundances 
used equivalent widths. The distortion of the line profiles could 
have, at most, increased the microturbulence velocity. It is known that Am 
stars show also distorted line profiles that lead to an increase in 
\vmic\ of 1.5-2.0\,\kms, but for Am stars the distortion is of the order of 
3-4\,\kms\ \citep{john1998}. If granulation played a role in increasing the 
derived value of \vmic, it is probably less than 0.1\,\kms, which 
is below our  detection limit.    

\begin{figure}[ht]
\begin{center}
\includegraphics[width=\hsize,clip]{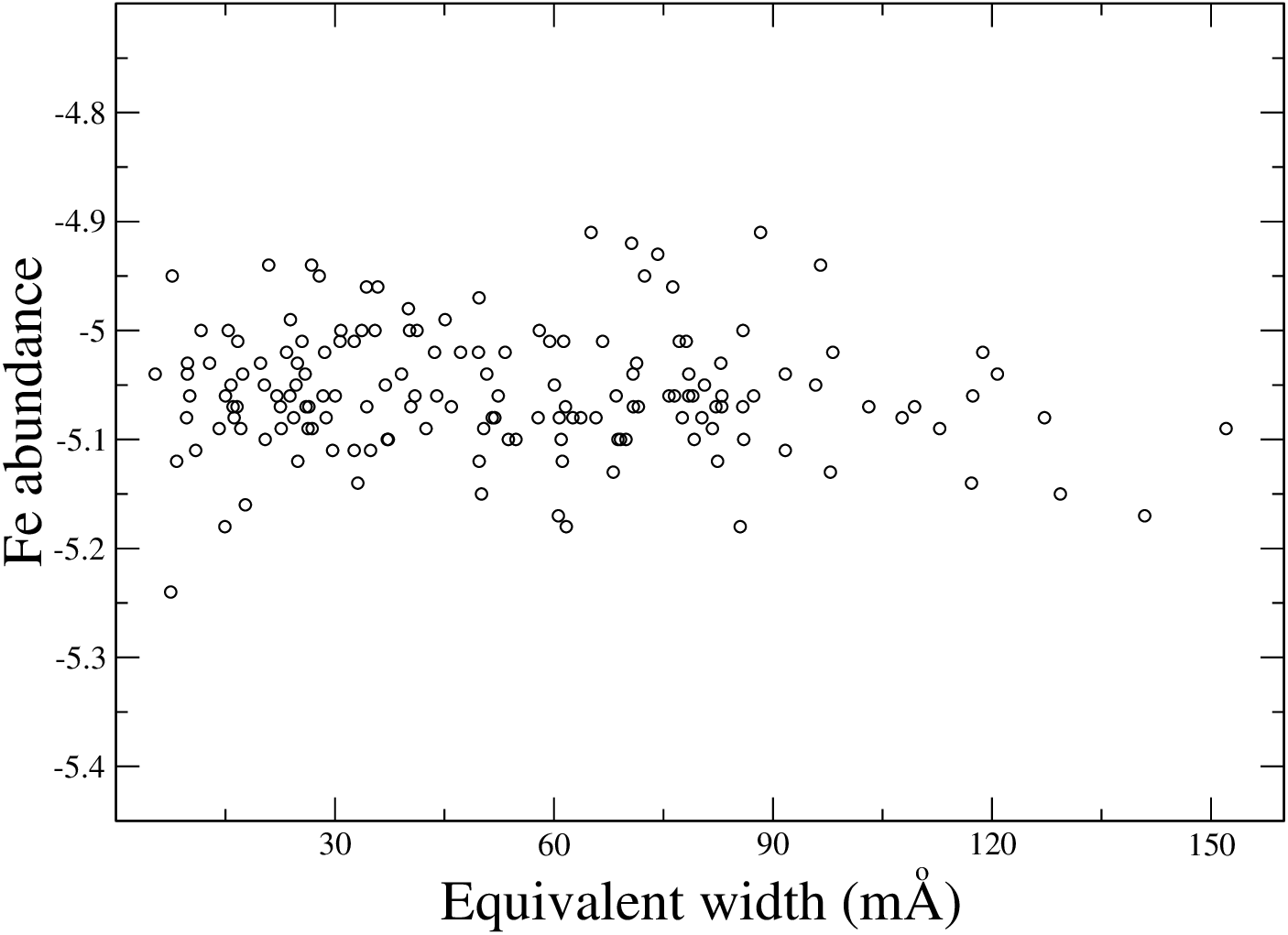}
\caption{\ion{Fe}{i} individual line abundances vs. equivalent widths. 
Abundances are derived for the preferred model parameters, \Teff=6500\,K,
\logg=4.00, \vmic=1.5\kms.}
\label{vmic}
\end{center}
\end{figure}

The microturbulence was determined by minimising the correlation between
equivalent width and abundance for several ions. We mainly used
\ion{Fe}{i} lines since this is the ion that provides the largest number 
of measured lines within a wide range in equivalent width, but also the
correlations obtained with \ion{Ti}{i}, \ion{Ti}{ii}, \ion{Cr}{i}, 
\ion{Fe}{ii}, and \ion{Ni}{i} lines were taken into account. The range 
of the microturbulent velocities goes from 1.4\,\kms\  (\ion{Ti}{i}) to 
1.9\,\kms\  (\ion{Ti}{ii}) with an average of \vmic=1.60 $\pm$ 0.18\,\kms.
The microturbulent velocity derived from \ion{Fe}{i} lines is 1.5\,\kms. This 
value, with an uncertainty of $0.2$\,\kms, was adopted for the final 
analysis as the best representation for all lines of neutral elements.  
Figure~\ref{vmic} (Online material) displays the correlation between the line 
abundance of \ion{Fe}{i} and the measured equivalent widths. 

The final abundances are given in Table~\ref{abundance}. We also computed 
the abundance difference between HD~49933 and the solar atmosphere as 
derived by \citet{met05} (4th column) and by \citet{sun_GS} (5th column). The 
results of the abundance analysis of HD~49933 taken from \citet{bruntt} are 
given in the 6th column of Table~\ref{abundance} for comparison.

The stellar metallicity ($Z$) is defined as follows:
\begin{equation}
\label{Z}
Z_{\rm star}=\frac{\sum_{a \geq 3}m_{a}10^{\log(N_{a}/N_{tot})}}{\sum_{a \geq 1}m_{a}10^{\log(N_{a}/N_{tot})}},
\end{equation}
where $a$ is the atomic number of an element with atomic mass m$_{\rm a}$.
Making use of the abundances obtained from the performed analysis we derived a
metallicity of $Z$ = 0.008 $\pm$ 0.002\,dex, adopting the solar abundances by 
\citet{met05} for all the elements that were not analysed. However, if we
assume an underabundance of $-0.5$\,dex for \textit{all} not analysed elements 
the resulting $Z$ value remains practically unchanged. Scaling the solar 
abundances of all elements by $-0.5$\,dex gives the metallicity 
$Z$ = 0.006\,dex. This substantial difference in $Z$ illustrates that 
it is important to have an accurate determination of C and O, which make a 
large contribution to the determination of $Z$. Additionally, a He 
underabundance of $-1$ and $-3$\,dex relative to Sun does not change
the $Z$ value.

\subsection{Abundance uncertainties}
The abundance uncertainties shown in Table~\ref{abundance} are the 
standard deviation of the mean abundance from the individual line abundances 
and do not represent the real error bars associated with each derived 
element abundance.

Based on realistic errors on the equivalent width measurements we 
find that the uncertainty on the abundances of a given line is only about 
0.01\,dex \citep[see ][ for more details]{fossati2009}.
In the case of ions with a sufficiently high number of lines we assume that the
internal scatter for each ion includes the uncertainties due to
equivalent width measurement and continuum normalisation.

Plotting the abundance scatter as a function of the number of lines it is
possible to conclude about the mean scatter that can be applied in cases
when only one line of a certain ion is measured. In these cases it is 
reasonable to assume an internal error of 0.10\,dex.

The internal scatter is just a part of the total abundance error bar since the
uncertainties on the fundamental parameters are also playing an important role.
Table~\ref{error} (Online material) shows the variation in abundance resulting 
from the changing of each parameter (\Teff, \logg\ and \vmic) by $+1\sigma$ 
and the final adopted uncertainty for each ion following the standard 
error analysis. 

The main source of the uncertainty is the effective temperature, 
while the uncertainties due to \logg\ and in particular to \vmic\ 
are almost negligible. 

Assuming that the contributions to the uncertainty on the abundances are 
independent, we derive the final uncertainty using error propagation. 
The result is given in column seven of Table~\ref{error} (available as 
Online material). 

\section{Discussion}\label{discussion}

\subsection{Comparison with previous determinations}\label{comp_param}
Our abundance calculation with the use of different atmospheric models for
HD~49933 proposed in literature shows that the accuracy (rms) of the 
abundance determination is the same for practically all of them and any 
difference in absolute value is caused by the differences in the model 
parameters. Therefore we shall compare and discuss here the 
determination of the atmospheric parameters.

The published effective temperature values range from 6300\,K 
\citep{hartmann} through 6484\,K \citep{blackwell} (IR flux method), 6595\,K 
\citep{edvardsson} (Str\"omgren photometry) to 6735--6780\,K 
\citep{gillon,bruntt04,bruntt}. The last three cited works employed spectrum 
synthesis iterative methods: APASS, VWA. For more details on previously 
derived atmospheric parameters see also Table~4 of \citet{bruntt04}. The APASS 
and VWA methods use the correlation of the abundance of neutral Fe lines with 
excitation potential as an indicator of \Teff. The final \Teff\ is found by 
minimising this correlation. Hydrogen lines are not considered in APASS and 
VWA methods. It is important to emphasise that in the APASS method 
\citep{gillon} the continuum level is adjusted in each iteration. 
\citet{gillon} also obtained very different temperature estimates working with 
APASS (\Teff=6735$\pm$53\,K) and with equivalent widths (\Teff=6538$\pm$44\,K). 
The authors gave a favor for the APASS results motivating that the Gaussian 
approximation of the line profiles (used to measure equivalent widths) in 
spectrum of a star rotating with \vsini=10\,\kms\ overestimates the equivalent 
widths. Since we used equivalent widths we investigated this possible 
error source by calculating  synthetic profiles of 26 \ion{Fe}{i} lines in 
the range 6--122\,m\AA\ with our adopted model atmosphere and with Fe 
abundance of $\log ({\rm Fe}/N_{\rm tot})=-5.05$, measured their equivalent 
widths with direct integration, broadened the synthetic spectrum with the 
instrumental, rotational and macroturbulent velocity profiles, and then again 
measured equivalent widths by Gaussian approximation of the convolved 
spectrum. Inserting both sets of equivalent widths in the \width\ code we 
got an average abundance of $\log ({\rm Fe}/N_{\rm tot})=-5.050\pm0.004$ 
adopting directly integrated equivalent widths and of 
$\log ({\rm Fe}/N_{\rm tot})=-5.040\pm0.016$  adopting the Gaussian 
approximation, as shown in Table~\ref{EqW} present in the Online material. No 
significant dependencies indicating any change of microturbulent velocity or 
effective temperature appeared in the second case. Thus we conclude that in 
HD~49933 the use of equivalent widths instead of synthetic spectrum should not 
influence the determination of the model parameters. 

\begin{figure*}[ht]
\begin{center}
\includegraphics[width=0.49\hsize,clip]{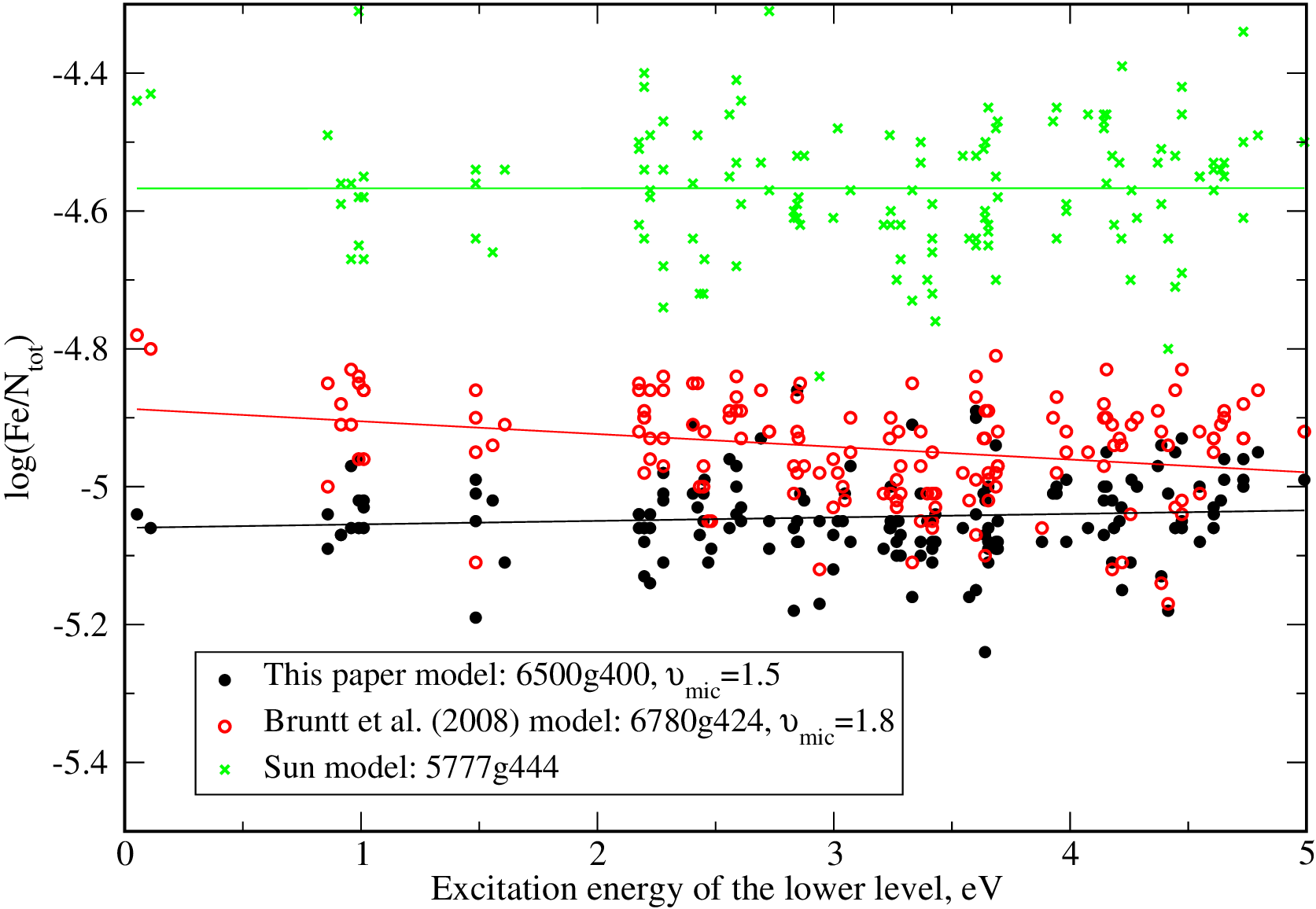}
\hspace{0.1cm}
\includegraphics[width=0.49\hsize,clip]{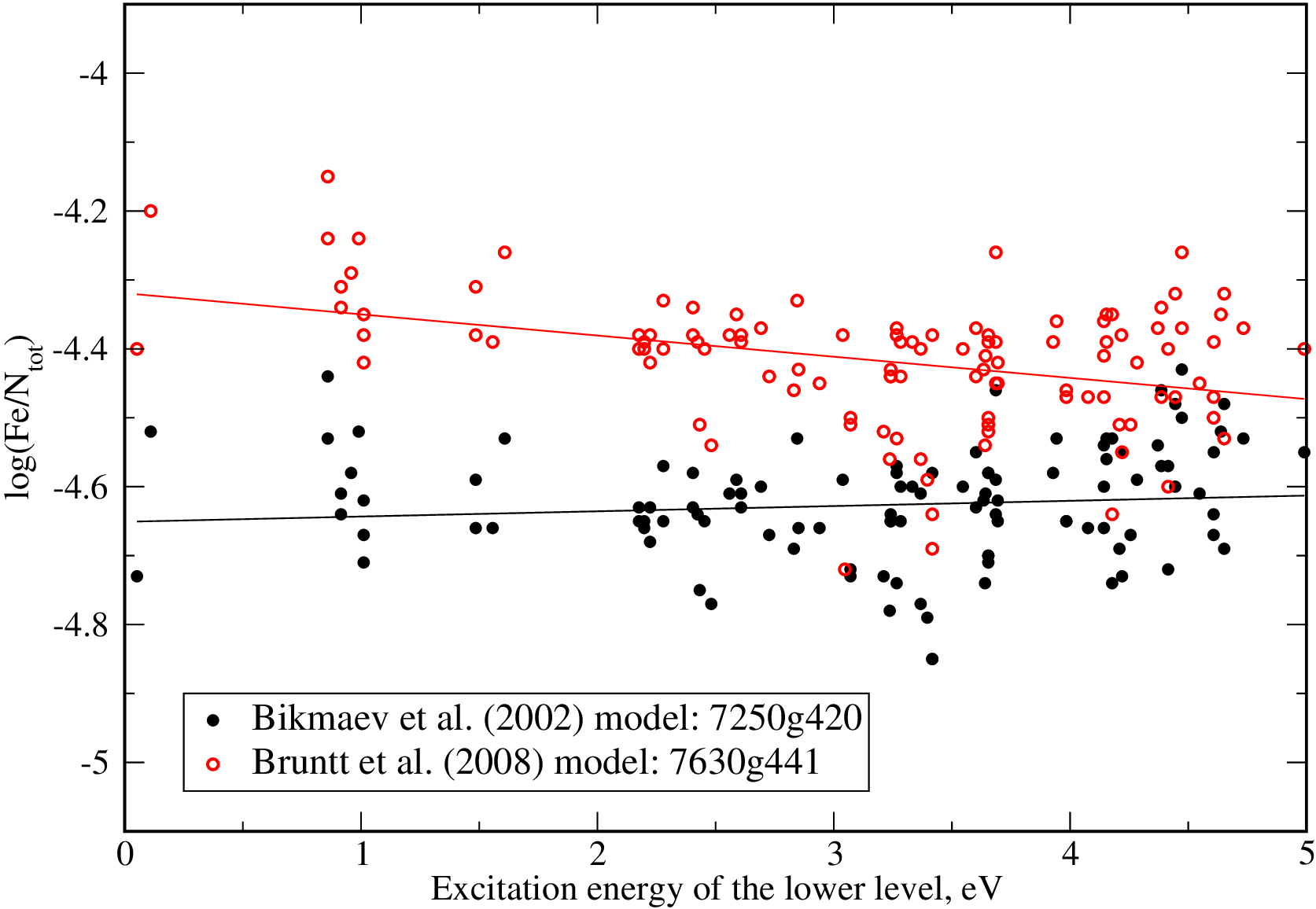}
\caption{Plots of individual abundances for 158 \ion{Fe} lines versus the 
excitation energy of the lower level for HD~49933 (left panel) and for 
HD~32115 (right panel) with different model atmospheres. The same dependence
for 137 common \ion{Fe} lines in solar spectrum is shown by crosses in the 
left panel.}
\label{Ab-Ei}
\end{center}
\end{figure*}

The use of both the APASS and the VWA \citep{bruntt} methods 
also leads to a higher \Teff\  for HD~49933 and another star HD~32115, 
for which the effective temperature was derived by \citet{bikmaev02} with the 
same method (Str\"omgren photometry and hydrogen line profiles) as in the 
present paper. Figure~\ref{Ab-Ei} shows individual abundances from 158 
\ion{Fe}{i} lines calculated with the model parameters derived in this paper
(\Teff\ = 6500\,K, \logg\ = 4.00) and by \citet{bruntt} 
(\Teff\ = 6780\,K, \logg\ = 4.24) versus the excitation energy of the lower 
level. For both models the microturbulent velocity, derived by us in the 
usual way, is nearly the same, 1.5\,\kms\ and 1.6\,\kms, respectively. However, 
to be consistent with the results by \citet{bruntt} we plot abundances versus 
excitation potential for Bruntt's model using \vmic=1.8\,\kms\ derived in 
\citet{bruntt}. The same dependence was also calculated for HD~32115 using 
equivalent width measurements from \citet{bikmaev02} and model atmospheres 
from \citet{bikmaev02} and from \citet{bruntt}, respectively.
As a final check we used the solar flux spectrum. We analysed 137 out of 
158 \ion{Fe}{i} lines with the solar model atmosphere (\Teff\ = 5777\,K, 
\logg\ = 4.44) also calculated with the \llm\ code. There were 21 
discarded lines that could not be properly fitted in the solar spectrum. 
Seven lines are blended in the Sun which is cooler and more metallic than 
HD~49933, while the rest 14 lines have extended wings and cannot be accurately 
measured. Our analysis resulted in $\log ({\rm Fe}/N_{\rm tot})=-4.57\pm0.09$ for 
\vmic=0.95\,\kms\ in full agreement with the currently adopted solar 
parameters \citep{met05}. Individual abundances derived from the solar 
\ion{Fe}{i} lines versus the excitation energy of the lower level are 
shown on Figure~\ref{Ab-Ei}.

In Fig.~\ref{Ab-Ei} one immediately notices the negative slope, indicating a 
too high \Teff, adopting the model parameters obtained by \citep{bruntt} for 
both stars. This fact together with the poor fit of the hydrogen line profiles 
(see Fig.~\ref{hydrogen}) clearly shows that both automatic methods 
based on synthetic spectrum calculations, ignoring hydrogen lines, lead to 
an overestimate of the effective temperature by $\sim$200\,K. With our model 
we get small positive slope that formally corresponds to an underestimate 
of the \Teff\ by $\sim$30\,K. For the Sun zero slope is obtained. 

Our temperature is close to that derived with the IR flux method by 
\citet{blackwell}, \Teff=6484$\pm$45\,K, and provides a good description 
of all global observables: photometric colors, hydrogen and metallic line 
profiles.

Our value for the surface gravity is supported by the ionisation equilibrium 
for \ion{Fe}{i}/\ion{Fe}{ii} and few other elements, such as 
\ion{Si}{i}/\ion{Si}{ii} and \ion{Ca}{i}/\ion{Ca}{ii}. 
Even for Ti and Cr, for which we have enough lines for an accurate 
abundance analysis, the average abundances from the two ionisation stages 
agree within the error bars. Applying a higher microturbulence of 1.9\,\kms,
derived from \ion{Ti}{ii} lines, for example, leads to a
\ion{Ti}{i}/\ion{Ti}{ii} equilibrium. In the case of chromium about half of 
the analysed \ion{Cr}{ii} lines and one third of \ion{Cr}{i} lines have 
theoretically calculated oscillator strengths which may influence the final 
abundance results. The atmospheric parameters derived for HD~49933 are 
similar to another solar-type pulsator, 
Procyon, which has \Teff=6512$\pm$49\,K, \logg=3.96$\pm$0.02, and 
$\log ({\rm Fe}/N_{\rm tot})=-4.60\pm0.15$  \citep{Procyon}. The 
comparison of the observed and the synthetic \ion{Mg}{i} line profile for the 
5172 and 5183\,\AA\ \ion{Mg}{i} lines for Procyon is shown in 
Fig.~\ref{mglines} \citep[see ][ for details concerning the observed spectrum of Procyon and the adopted model atmosphere]{tanya-reference}.
\subsection{Synthetic photometry}\label{synth-phot}
Since there are no available observations of HD~49933 in the visible spectral
region, calibrated to physical units, we adopted photometric observations 
extracted from the SIMBAD 
database\footnote{{\tt http://simbad.u-strasbg.fr/simbad/}}. 
Table~\ref{tab:colors} summarises the comparison between observed and 
theoretical color-indexes of different photometric systems. The
theoretical colors were computed using computer codes by 
\citet{kurucz1993a} modified to read and process high-resolution fluxes 
produced by the \llm. The reddening, corresponding to $E(B-V)=0$, was derived 
using analytical extinction models by \citet{ebv}.
Besides the final model with \Teff$=6500$\,K, \logg$=4.0$ (Model~1), we 
present also the synthetic photometry for three other models: 
\Teff$=6500$\,K, \logg$=3.85$ (Model~2); a model with the parameters 
derived from the Str\"omgren photometry (\Teff$=6550$\,K, \logg$=4.25$, 
Model~3), and a model with the parameters published by \citet{bruntt} 
(\Teff$=6780$\,K, \logg$=4.24$, Model~4).
\begin{table}
\caption{Observed and calculated photometric parameters of HD~49933. The values in brackets give the error bars 
of observations.}
\begin{scriptsize}
\begin{center}
\begin{tabular}{cccccccc}
\hline\hline
Color           & \textbf{SIMBAD}   & t6500g4.0 & t6500g3.85  & t6550g4.25 & t6780g4.24 \\
index           &                   & (Model~1) & (Model~2)   & (Model~3)  & (Model~4) \\
\hline
$b$-$y$         &$\mathbf{0.270}$   &$0.2754$   &$0.2727$	  &$0.2724$    &$0.2399$     \\ 
                & ($0.002$) &&&\\
$m_{\rm 1}$     &$\mathbf{0.127}$   &$0.1251$   &$0.1255$	  &$0.1330$    &$0.1355$     \\ 
                & ($0.004$) &&&\\
$c_{\rm 1}$     &$\mathbf{0.460}$   &$0.5123$   &$0.5502$	  &$0.4400$    &$0.5126$     \\ 
                & ($0.003$) &&&\\
$H\beta$        &$\mathbf{2.662}$   &$2.7272$   &$2.7291$	  &$2.7279$    &$2.7501$     \\ 
$B$-$V$         &$\mathbf{0.390}$   &$0.3865$   &$0.3829$	  &$0.3865$    &$0.3371$     \\ 
$U$-$B$         &$\mathbf{-0.070}$  &$-0.0628$  &$-0.0457$	  &$-0.0942$   &$-0.0882$   \\
\\
\textbf{Geneva} &&&&\\
$U$-$B$         &$\mathbf{1.235}$   &$1.2561$   &$1.2896$	  &$1.1950$    &$1.2285$     \\ 
$V$-$B$         &$\mathbf{0.491}$   &$0.5125$   &$0.5166$	  &$0.5123$    &$0.5700$     \\ 
$B_{\rm1}$-$B$  &$\mathbf{0.973}$   &$0.9842$   &$0.9835$	  &$0.9878$    &$0.9723$     \\ 
$B_{\rm2}$-$B$  &$\mathbf{1.391}$   &$1.4142$   &$1.4152$	  &$1.4102$    &$1.4252$     \\ 
$V_{\rm1}$-$B$  &$\mathbf{1.224}$   &$1.2423$   &$1.2461$	  &$1.2420$    &$1.2962$     \\ 
$G$-$B$         &$\mathbf{1.586}$   &$1.6135$   &$1.6185$	  &$1.6146$    &$1.6847$     \\ 
\hline                                                                                
\end{tabular}                                                                         
\end{center}
\label{tab:colors}
\end{scriptsize}
\end{table}

%
It is seen from the Table~\ref{tab:colors} that the Str\"omgren and UBV 
indexes are better described by Model~1. 
The H$\beta$ index shows a poor fit for all models, but this does not 
play a critical role in the present study since the profiles of hydrogen lines 
were fitted perfectly with the parameters of Model~1. The color-indexes of the 
Geneva system are generally better described by Model~1,2 and 3
i.e. models with lower \Teff. However, the indexes $U-B$ and $B_{\rm 1}-B$ are 
better fitted with a higher temperature (Model~4). Model 1 and 3 show a
comparable fit, with the difference that Model 1 fits better the \logg\ 
sensitive $c1$ index, but worse the $U-B$ index, while for Model 3 it is the
opposite. In summary, we find that the majority of the photometric 
indicators and the present accurate spectroscopic study point to the adopted 
\Teff$=6500$\,K and \logg=4.0. 

\section{Conclusions}
Based on ten high-resolution, high signal-to-noise spectra taken from 
ESO archive we carried out a precise spectroscopic analysis of one of the
primary solar-like CoRoT targets HD~49933. All ten spectra were averaged 
to obtain a single spectrum with a resolution $R\approx115000$ and 
signal-to-noise ratio $\approx500$ per pixel at $\lambda\approx5000$\AA.
Using this spectrum we revised the fundamental parameters and the atmospheric 
abundances in a consistent way employing modern 1D LTE stellar model 
atmospheres \citep[LLmodels; ][]{llm}. The derived set of fundamental 
parameters and abundances provide a good fit to the available 
observables: IR and multicolor photometry, pressure-sensitive magnesium 
lines, metallic lines and profiles of hydrogen Balmer lines.

It is shown that the implementation of automatic procedures for the abundance 
analysis using only line spectra without hydrogen lines can result in
inaccurate parameters and hence abundances.

\begin{acknowledgements}
This work was supported by following funding projects: FWF Lise Meitner grant 
Nr. M998-N16 (DS), by the Presidium RAS Programme ``Origin and evolution
of stars and galaxies'', and by the Leading Scientific School grant 
4354.2008.2 (TR). LF has received support from the Austrian Science Foundation 
(FWF project P19962 - Modulated RR Lyrae Stars). We kindly thank Thomas 
Kallinger and Michael Gruberbauer for the useful discussions and comments 
during the preparation of the draft.
\end{acknowledgements}
\Online
%
\begin{table*}[ht]
\caption{
Error sources for the abundances of the chemical elements in HD~49933.
Column 3 gives the standard deviation $\sigma_{\rm abn}$\,(scatt.) 
of the mean abundance obtained
from different spectral lines (internal scatter); a blank means that the number
of spectral lines is $<2$, hence no internal scatter could be estimated.
(Note that these values are identical to those given in Table~\ref{abundance}.)
Columns 4, 5, and 6 give the variation in abundance estimated by increasing
\Teff\ by 50\,K, \logg\ by 0.15\,dex, and \vmic\ by 0.2\,km\,s$^{-1}$,
respectively. Column~7 gives the the mean error calculated applying
standard error propagation theory on the uncertainties given in the
previous columns, i.e.,
$\sigma_{\rm abn}^2$\,(tot)    = 
$\sigma_{\rm abn}^2$\,(scatt.) + 
$\sigma_{\rm abn}^2$\,(\Teff)  + 
$\sigma_{\rm abn}^2$\,(\logg)  + 
$\sigma_{\rm abn}^2$\,(\vmic).
For the computation of $\sigma_{\rm abn}^2$\,(tot) of those ions for which
the internal scatter could not be measured, we have assumed \textit{a priori} 
$\sigma_{\rm abn}\,{\rm (scatt.)} = 0.10$\,dex.
}
\label{error}
\begin{center}
\begin{tabular}{lrrrrrr}
\hline
\hline
\multicolumn{1}{c}{Ion         }        &  
\multicolumn{1}{c}{abundance   }        &  
\multicolumn{1}{c}{$\sigma_{\rm abn}$\,(scatt.)}  &  
\multicolumn{1}{c}{$\sigma_{\rm abn}$\,(\Teff) }  &  
\multicolumn{1}{c}{$\sigma_{\rm abn}$\,(\logg) }  &  
\multicolumn{1}{c}{$\sigma_{\rm abn}$\,(\vmic) }  &  
\multicolumn{1}{c}{$\sigma_{\rm abn}$\,(tot)   }  \\ 
\multicolumn{1}{c}{            }        & 
\multicolumn{1}{c}{$\log (N/N_{\rm tot})$} & 
\multicolumn{1}{c}{(dex)       }        & 
\multicolumn{1}{c}{(dex)       }        & 
\multicolumn{1}{c}{(dex)       }        & 
\multicolumn{1}{c}{(dex)       }        & 
\multicolumn{1}{c}{(dex)       }        \\ 
\hline
\ion{C}{i}    & $-$3.74 & 0.10 & $-$0.02 & $ $0.05 & $ $0.00 & 0.11 \\
\ion{O}{i}    & $-$3.55 &      & $-$0.03 & $ $0.05 & $ $0.00 & 0.12 \\
\ion{Na}{i}   & $-$6.15 & 0.05 & $ $0.02 & $ $0.00 & $-$0.01 & 0.05 \\
\ion{Mg}{i}   & $-$4.83 & 0.07 & $ $0.03 & $-$0.01 & $-$0.02 & 0.08 \\
\ion{Mg}{ii}  & $-$4.73 &      & $-$0.02 & $ $0.05 & $-$0.01 & 0.11 \\
\ion{Al}{i}   & $-$6.20 &      & $ $0.02 & $ $0.00 & $ $0.00 & 0.10 \\
\ion{Si}{i}   & $-$4.86 & 0.21 & $ $0.01 & $ $0.00 & $-$0.01 & 0.21 \\
\ion{Si}{ii}  & $-$4.82 & 0.02 & $-$0.03 & $ $0.05 & $-$0.02 & 0.07 \\
\ion{S}{i}    & $-$5.23 & 0.07 & $-$0.01 & $ $0.04 & $ $0.00 & 0.08 \\
\ion{Ca}{i}   & $-$6.01 & 0.11 & $ $0.03 & $-$0.01 & $-$0.05 & 0.12 \\
\ion{Ca}{ii}  & $-$6.01 & 0.09 & $-$0.02 & $ $0.05 & $-$0.01 & 0.11 \\
\ion{Sc}{ii}  & $-$9.24 & 0.12 & $ $0.02 & $ $0.05 & $-$0.05 & 0.14 \\
\ion{Ti}{i}   & $-$7.54 & 0.07 & $ $0.03 & $ $0.00 & $-$0.02 & 0.08 \\
\ion{Ti}{ii}  & $-$7.42 & 0.12 & $ $0.01 & $ $0.05 & $-$0.05 & 0.16 \\
\ion{V}{i}    & $-$8.50 & 0.13 & $ $0.04 & $ $0.00 & $-$0.01 & 0.15 \\
\ion{V}{ii}   & $-$8.47 & 0.23 & $ $0.01 & $ $0.05 & $-$0.01 & 0.15 \\
\ion{Cr}{i}   & $-$6.82 & 0.17 & $ $0.04 & $ $0.00 & $-$0.02 & 0.18 \\
\ion{Cr}{ii}  & $-$6.61 & 0.17 & $ $0.00 & $ $0.05 & $-$0.03 & 0.19 \\
\ion{Mn}{i}   & $-$7.33 & 0.14 & $ $0.03 & $ $0.00 & $-$0.03 & 0.15 \\
\ion{Fe}{i}   & $-$5.04 & 0.06 & $ $0.04 & $-$0.01 & $-$0.04 & 0.13 \\
\ion{Fe}{ii}  & $-$5.03 & 0.08 & $ $0.01 & $ $0.05 & $-$0.05 & 0.12 \\
\ion{Co}{i}   & $-$7.49 & 0.10 & $ $0.03 & $ $0.00 & $ $0.00 & 0.10 \\
\ion{Ni}{i}   & $-$6.34 & 0.10 & $ $0.03 & $ $0.00 & $-$0.02 & 0.11 \\
\ion{Cu}{i}   & $-$8.65 & 0.07 & $ $0.03 & $ $0.00 & $-$0.01 & 0.08 \\
\ion{Zn}{i}   & $-$8.12 & 0.06 & $ $0.02 & $ $0.01 & $-$0.04 & 0.08 \\
\ion{Sr}{i}   & $-$9.65 &      & $ $0.03 & $ $0.00 & $ $0.00 & 0.10 \\
\ion{Sr}{ii}  & $-$9.50 & 0.04 & $ $0.01 & $ $0.06 & $-$0.01 & 0.07 \\
\ion{Y}{ii}   & $-$10.34& 0.10 & $ $0.02 & $ $0.05 & $-$0.02 & 0.12 \\
\ion{Zr}{ii}  & $-$9.85 & 0.06 & $ $0.01 & $ $0.05 & $-$0.01 & 0.08 \\
\ion{Ba}{ii}  & $-$10.06& 0.19 & $ $0.03 & $ $0.02 & $-$0.07 & 0.16 \\
\ion{La}{ii}  & $-$11.21& 0.11 & $ $0.03 & $ $0.06 & $ $0.00 & 0.14 \\
\ion{Ce}{ii}  & $-$10.73& 0.10 & $ $0.03 & $ $0.05 & $ $0.00 & 0.12 \\
\ion{Nd}{ii}  & $-$10.77& 0.28 & $ $0.02 & $ $0.05 & $-$0.01 & 0.28 \\
\ion{Sm}{ii}  & $-$11.09& 0.16 & $ $0.03 & $ $0.05 & $ $0.00 & 0.17 \\
\ion{Eu}{ii}  & $-$11.92& 0.10 & $ $0.03 & $ $0.05 & $ $0.00 & 0.12 \\
\ion{Gd}{ii}  & $-$11.16& 0.09 & $ $0.03 & $ $0.05 & $ $0.00 & 0.15 \\
\ion{Dy}{ii}  & $-$11.36& 0.15 & $ $0.03 & $ $0.05 & $ $0.00 & 0.16 \\
\hline								
\end{tabular}
\end{center}
\smallskip
\end{table*}

\begin{table*}[ht]
\caption{Comparison between the line abundances obtained for a set of \ion{Fe}{i} lines
calculated from equivalent widths measured with direct integration and a Gaussian approximation.}
\label{EqW}
\begin{center}
\begin{tabular}{lcccc}
\hline
\hline
 & \multicolumn{2}{c}{Direct integration} & \multicolumn{2}{c}{Gaussian measurements} \\
Wavelength & EqW  & Abundance & EqW  & Abundance \\
 \AA	   & m\AA &           & m\AA &           \\
\hline
5123.7200 &  60.30 & $-$5.05 &  61.31 & $-$5.02 \\
5127.3593 &  52.50 & $-$5.05 &  53.41 & $-$5.03 \\
5133.6885 &  87.60 & $-$5.06 &  87.23 & $-$5.06 \\
5139.4628 & 105.10 & $-$5.05 & 104.48 & $-$5.06 \\
5141.7393 &  53.20 & $-$5.05 &  54.11 & $-$5.03 \\
5142.9285 &  62.00 & $-$5.05 &  63.03 & $-$5.02 \\
5162.2729 &  85.60 & $-$5.05 &  84.42 & $-$5.07 \\
5165.4100 &  73.90 & $-$5.05 &  74.55 & $-$5.04 \\
5171.5964 &  98.20 & $-$5.05 &  99.20 & $-$5.03 \\
5184.2661 &  22.80 & $-$5.05 &  23.24 & $-$5.04 \\
5191.4550 &  98.80 & $-$5.05 &  98.57 & $-$5.05 \\
5192.3442 & 107.70 & $-$5.05 & 106.98 & $-$5.06 \\
5194.9418 &  82.90 & $-$5.05 &  84.00 & $-$5.02 \\
5602.9451 &  67.20 & $-$5.05 &  68.04 & $-$5.03 \\
5615.2966 &  24.00 & $-$5.05 &  24.30 & $-$5.04 \\
5615.6439 & 122.60 & $-$5.06 & 120.95 & $-$5.08 \\
5618.6327 &  15.70 & $-$5.05 &  16.10 & $-$5.03 \\
5620.4924 &   6.10 & $-$5.05 &   6.24 & $-$5.04 \\
5624.0220 &   7.80 & $-$5.04 &   7.74 & $-$5.05 \\
5624.5422 &  72.50 & $-$5.05 &  73.44 & $-$5.03 \\
5631.7310 &   6.30 & $-$5.04 &   6.39 & $-$5.04 \\
5633.9465 &  28.50 & $-$5.05 &  29.03 & $-$5.03 \\
5638.2621 &  31.00 & $-$5.05 &  31.58 & $-$5.03 \\
5640.3070 &   6.20 & $-$5.05 &   6.33 & $-$5.04 \\
5641.4340 &  17.50 & $-$5.05 &  17.88 & $-$5.03 \\
5655.4930 &  32.60 & $-$5.05 &  33.20 & $-$5.03 \\
\hline		        	       
Mean $\pm$ rms & & -5.050$\pm$0.004 &  & -5.040$\pm$0.016 \\								
\hline		        	       
\end{tabular}	        	       
\end{center}	        	       
\smallskip	        	       
\end{table*}

%
%
%
\end{document}